\documentclass[final,3p]{elsarticle}
\journal{Annals of Physics}
\usepackage{graphicx}
\usepackage{amsmath,amsfonts,amssymb,amsthm}
\usepackage{color}
\usepackage{soul}
\usepackage{subcaption}
\usepackage{caption}
\usepackage{physics}
\usepackage{dsfont}
\usepackage{hyperref}
\usepackage{url}
\usepackage{leftidx}
\usepackage{textcomp}
\usepackage{gensymb}
\usepackage{threeparttable}
\usepackage{rotating}
\usepackage{esvect}
\usepackage{booktabs}
\usepackage{marvosym}
\usepackage{lmodern}
\usepackage[ruled,vlined]{algorithm2e}
\usepackage[noend]{algpseudocode}
\usepackage[dvipsnames]{xcolor}
\usepackage{mathrsfs}
\usepackage{ytableau}
\usepackage{mathtools}

\definecolor{emerald}{rgb}{0.31, 0.78, 0.47}
\definecolor{blue(ncs)}{rgb}{0.0, 0.53, 0.74}

\hypersetup{
     colorlinks=true,
     linkcolor=magenta,
     filecolor=blue,
     citecolor=emerald,      
     urlcolor =blue(ncs),
}

\usepackage{chngcntr}

\DeclareMathAlphabet{\pazocal}{OMS}{zplm}{m}{n}

\newcommand{\twosig}{\includegraphics[trim=0 5pt 0 0, scale=1]{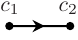}}
\newcommand{\ntwosig}{\includegraphics[trim=0 5pt 0 0, scale=1]{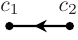}}
\newcommand{\threesig}{\includegraphics[trim=0 13pt 0 0, scale=1]{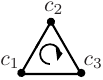}}
\newcommand{\nthreesig}{\includegraphics[trim=0 13pt 0 0, scale=1]{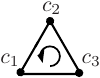}}
\newcommand{\foursig}{\includegraphics[trim=0 28pt 0 0, scale=1]{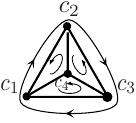}}
\newcommand{\nfoursig}{\includegraphics[trim=0 28pt 0 0, scale=1]{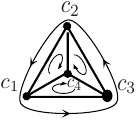}}

\newcommand{\AB}{\includegraphics[trim=0 13pt 0 0, scale=1]{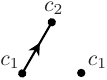}}
\newcommand{\BC}{\includegraphics[trim=0 13pt 0 0, scale=1]{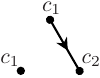}}
\newcommand{\AC}{\includegraphics[trim=0 15pt 0 0, scale=1]{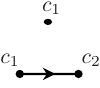}}

\newcommand{\foursite}{\includegraphics[trim=0 13pt 0 0, scale=1]{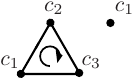}}
\newcommand{\foursiteA}{\includegraphics[trim=0 16pt 0 0, scale=1]{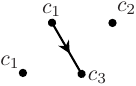}}
\newcommand{\foursiteB}{\includegraphics[trim=0 16pt 0 0, scale=1]{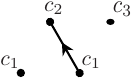}}
\newcommand{\fivesite}{\includegraphics[trim=0 28pt 0 0, scale=1]{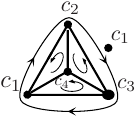}}
\newcommand{\fivesiteA}{\includegraphics[trim=0 16pt 0 0, scale=1]{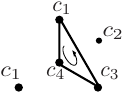}}
\newcommand{\fivesiteB}{\includegraphics[trim=0 16pt 0 0, scale=1]{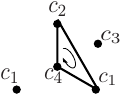}}
\newcommand{\fivesiteC}{\includegraphics[trim=0 16pt 0 0, scale=1]{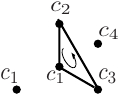}}

\begin{document}
\begin{frontmatter}

\title{Entanglement blossom in a simplex matryoshka}
  
\author{Zhao Zhang}
\ead{zz8py@virginia.edu}
 \affiliation{organization={SISSA and INFN, Sezione di Trieste},
	 addressline={via Bonomea 265},
	 postcode={I-34136}, 
	 city={Trieste}, 
	 country={Italy}}

\date{\today}

\begin{abstract}
Exotic entanglement entropy scaling properties usually present interesting entanglement structures in real space and novel metrics of the spacetime lattice. One prominent example is the rainbow chain where lattice sites which are symmetric about the center form entangled Bell pairs due to an effective long-range coupling in the strong inhomogeneity regime. This manuscript generalizes the rainbow chain to a Hausdorff dimension one lattice embedded in higher dimensional space and enlarged local Hilbert space keeping the Hamiltonian frustration free. The effective Hamiltonian from the Schrieffer-Wolf transformation is given by a stacking of layers of $k$-simplices with $0$-dimensional (all-to-all interacting) antiferromagnetic Hamiltonians, which can be diagonalized analytically with Young operators. The original lattice can be obtained by introducing disinclination defects in a regular $k$-dimensional cubical lattice, which introduces curvature at the center of the lattice. The model interpolates between the SYK model and the free-fermionic XX spin chain, and hence might be potentially useful in understanding black hole physics and holography.
\end{abstract}

\end{frontmatter}

%
\section{Introduction}
\label{sec:Intro}
%

Entanglement is an invaluable tool in understanding the structure of phases in many-body systems. In many cases we can visualize the entanglement is to think of the degrees of freedom as forming singlets of Bell pairs that carry units of entanglement entropy. When the singlets are between neighboring sites, such a picture leads to dimer and valence bond solid in the Majumdar-Ghosh~\cite{Majumdar:1969aa} and AKLT chain~\cite{PhysRevLett.59.799} in one dimensional systems with matrix-product ground states obeying the area law of entanglement entropy~\cite{Hastings:2007aa} for gapped systems. When the singlets can be formed between sites arbitrarily far away, as in the Motzkin and Fredkin spin chains, the system can go through an entanglement phase transition between area law and extensive scaling of entanglement~\cite{BravyiEA12,Movassagh13278,Salberger:2017aa,PhysRevB.94.155140,Zhang5142,Salberger_2017,Zhang_2017}, with ground states described by holographic tensor networks~\cite{PhysRevB.100.214430,Alexander2021exactholographic}. In two dimensions, this leads to Anderson's idea of resonating valance bonds~\cite{Anderson:1973aa,Fazekas:1974aa}, made concrete in the Rokhsar-Kivelson model~\cite{PhysRevLett.61.2376}. They belong to a more general class of vertex or tiling models with local constraints, where the singlet is formed between different local configurations of a lattice plaquette. The entanglement entropy of their ground state naturally obeys the area law, as a projected entangled pair state with the tensor network being the lattice itself~\cite{Zhang_2023}. Yet, by decorating such models with a color degree of freedom, it is possible to make the singlets formed by the coloring instead, while the vertex or tiling configurations facilitate them to be separated arbitrarily far away across the lattice~\cite{zhang2022entanglement,zhang2022area}. The average distance singlets span is again controlled by the local deformation parameter, resulting in a quantum phase transition between area-law and volumetric scaling of entanglement entropy between half systems cut in either direction.

All of the above mentioned models share the common feature of being frustration free, making it convenient to write down a unique ground state that allows exact results analytically. Frustration becomes an obstacle when generalizing such models to singlet states among more than two sites, either in the form of trimer, n-mer, valance bond solid~\cite{PhysRevB.75.060401,PhysRevB.75.184441} in one dimensional chains, or simplex solid states~\cite{PhysRevB.77.104404} in higher dimensions, unless the local Hibert space is enlarged to the corresponding dimension. Such extensions not only provide benchmark for relevant cold atom experiments~\cite{Gorshkov:2010aa,Taie:2012aa,Pagano:2014aa,Scazza:2014aa,Zhang_2014}, but also prove useful in the understanding of entanglement structure even when frustration is present~\cite{PhysRevX.4.011025}. In fact, the n-singlet picture turns out to be revealing in understanding the entanglement structure of multi-component generalizations to the plain Heisenberg~\cite{Sutherlandprb} and XXZ spin chains~\cite{mussardo2020prime, PhysRevB.106.134420} with permutation Hamiltonian, which is not deliberately cast into projection operators that are frustration free.

An alternative mechanism to generate long-range entangled singlet pairs is by introducing strong inhomogeneity in the XX spin chain. This was done in the so-called rainbow chain, as the singlets are at fixed locations pairwise symmetric about the center of the chain, giving a maximal entanglement between left and right half chains~\cite{Vitagliano_2010}. The inhomogeneity can be interpreted as an underlying spacetime with constant negative curvature in the continuous free-fermionic version of the model~\cite{Rodriguez-Laguna:2017aa}, and the conformal field theory that describes the free fermions has a holographic dual in the Anti-de Sitter space $AdS_2$~\cite{MacCormack:2019aa}. The metric allows more refined structure of the entanglement than the entropy to be computed, such as the entanglement Hamiltonian and the entanglement contour~\cite{Tonni:2018aa}. In Ref.~\cite{Ramirez:2015aa}, a first attempt at generalizing the mechanism to two dimension was made with an anisotropic quasi-two dimensional model inhomogeneous in one direction but translationally invariant in the other. 

In this manuscript, a straightforward isotropic generalization is realized in the spirit of simplex singlet state by enlarging the dimensionality of the local Hilbert space. Although the lattice lives in a two-dimensional manifold, its Hausdorff dimension is still one. However, the location of lattice sites sheds light upon an interpretation of the inhomogeneity of coupling strength as a natural result of its exponential decay over distance, which is absent in the 1D case. The rest of the paper is organized as follows. In Sec.~\ref{sec:flower}, the 2D generalization to rainbow chain is defined on the floral lattice, and the strong disorder renormalizaiton group (RG) procedure is carried out to show the effective Hamiltonian and its ground state. In Sec.~\ref{sec:curve}, the lattice geometry is briefly discussed to show the positive curvature near the center and how it can be obtained from square lattice by proliferating disclination defects. In Sec.~\ref{sec:matryoshka}, the model is further generalized to three and higher dimensions outlining the analogous RG transformation. Finally, a conclusion is given in Sec.~\ref{sec:Concl} with a few possible future directions.

%
\section{Inhomogeneous XX model on the floral lattice}
\label{sec:flower}
%
The key observation of generalizing the rainbow chain to higher dimensional space is that its lattice sites at the same radius from the center of the system form a maximally entangled singlet state. For a lattice to span a two dimensional space, there need to be at least three lattice sites within each layer of a certain radius. This in turn requires the local Hilbert space of each site to have three degrees of freedom, so that there exist a unique singlet state that antisymmetrizes the wave-function of the three-body system. Given that the rainbow chain in 1D is simply an inhomogeneous deformation to the free-fermion XX spin chain, one might be tempted to choose as the nearest neighbour Hamiltonian term that of the $SU(3)$ generalization to the XX spin chain, which is given in terms of the ladder operators corresponding to the two simple roots, which can be mapped to two species of free fermions~\cite{Mutter:1995aa, Maassarani:1998aa}. However, here we are dealing with a few-body problem with exact diagonalization at each order of the Dasgupta-Ma renormalization~\cite{PhysRevB.22.1305,Vitagliano_2010}, and the integrability of the Hamiltonian is not necessary. Moreover, to guarantee the Hamiltonian to be invariant in the RG procedure, it needs to respect the $S_3$ permutation symmetry by including all three pairs of ladder operators
\begin{equation}
	e^1=\begin{pmatrix}
		0 & 1 & 0 \\
		0 & 0 & 0 \\
		0 & 0 & 0 
	\end{pmatrix}, \quad
	e^2=\begin{pmatrix}
		0 & 0 & 0 \\
		0 & 0 & 1 \\
		0 & 0 & 0 
	\end{pmatrix}, \quad e^3=\begin{pmatrix}
	0 & 0 & 0 \\
	0 & 0 & 0 \\
	1 & 0 & 0 
	\end{pmatrix},
\end{equation}
and their conjugates $f^a={e^a}^\dagger$, $a=1,2,3$. The local Hamiltonian between neighboring site $i$ and $j$ is given by 
\begin{equation}
	h_{i,j}=\sum_{a=1}^3(e^a_i f^a_j+ f^a_i e^a_j)\equiv 2\sum_{\substack{a=1\\ a\ne 3}}^7 \lambda^a_i\lambda^a_j,
	\label{eq:offdiagh}
\end{equation}
where among the Gell-Mann generators of $SU(3)$, $\lambda_3$ and  $\lambda_8$ spanning the Cartan subalgebra are excluded in the sum.
We can denote the three components of the local Hilbert space $\mathbb{C}^3$ by colors red ($R$), green ($G$) and blue ($B$). The Hamiltonian consists of kinetic terms that exchange colors between neighboring pairs of different colors, and its lowest energy eigenstate is simply the antisymmetrization of whatever states the two neighboring sites are in.

\begin{figure}[t!bh]
	\centering
	\begin{subfigure}[b]{0.4\textwidth}
		\centering
		\includegraphics[width=\textwidth]{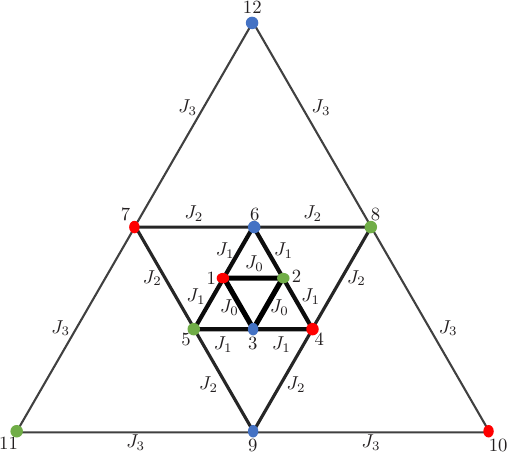} 
		\caption{}
		\label{fig:flower1}
	\end{subfigure}
	\begin{subfigure}[b]{0.4\textwidth}
		\centering
		\includegraphics[width=\textwidth]{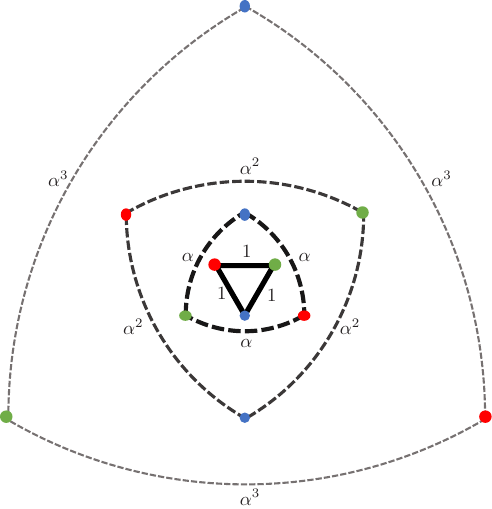}
		\caption{} 
		\label{fig:flower2}
	\end{subfigure}
	\caption{(a) Nested triangle lattice with inhomogeneous coupling strength between neighboring sites that decays geometrically with distance. (d) The effective long-range coupling between lattice cites with the same radius.}
	\label{fig:flower}
\end{figure}

Now that the form of the local interaction is fixed, the next question is how to choose the adjacency of the lattice, i.e. the interaction between which pairs should be turned on, and how the coupling strength should decay from layer to layer. The intuitive observation of the rainbow chain is that every time a new layer is added to the system, each new lattice site has  antiferromagnetic interaction with all the other sites in the previous layer but one, such that if the all the sites in 
the previous layer are already favored to be in different configurations, enforced by the previous order of perturbation, the same must go for all the sites in the current layer, which effectively results in a new singlet state. In two dimension, such a consideration leads to the lattice in Fig.~\ref{fig:flower1}.

We begin by considering a lattice of size 6, as shown in the two innermost layers in Fig.~\ref{fig:flower1}, with the Hamiltonian
\begin{align}
	H_1=&H_0+J_1V_1,\\
	H_0=&h_{1,2}+h_{2,3}+h_{3,1},\\
	V_1=&h_{3,4}+h_{4,2}+h_{1,5}+h_{5,3}+h_{2,6}+h_{6,1}.
\end{align}
Let $J_1=\sqrt{6}\alpha^{\frac{1}{2}}$, then in the case of $\alpha \ll 1$, it can be transformed into a low energy effective Hamiltonian in the ground state subspace of the zeroth order Hamiltonian $H_0$, with the standard Schrieffer-Wolff transformation~\cite{PhysRev.149.491,Bravyi:2011aa}. Since the number of sites is the same as the dimensionality of the local Hilbert space, $H_0$ is frustration-free, so its unique ground state is the simultaneous lowest energy eigenstate of each $h_{i,j}$, i.e.~the fully antisymmetrized state with energy $-3$,
\begin{equation}
	|p_0\rangle=|\mathcal{A}\rangle_{123}\equiv\frac{1}{\sqrt{6}}\sum_{\sigma\in S_3}\mathrm{sgn}(\sigma) |\sigma(RGB)\rangle_{123},
\end{equation}
where $\mathrm{sgn}(\sigma)$ denotes the signature of the permutation $\sigma$, and the configuration of site $i$ is labeled by the $i$'th letter denoting its color in the ket-vector\footnote{The diagonalization of $H_0$ is carried out in \ref{sec:Young}.}. Even without any explicit calculation, one can already see from the symmetry and the absence of diagonal terms of the Hamiltonian, that by construction, the effective Hamiltonian at the next order will be of the same form, up to an overall constant energy shift by the identity operator. In Sec.~\ref{sec:SW}, one matrix element of diagonal and off-diagonal terms are evaluated explicitly to fix the coefficients, giving 
\begin{equation}
	H^{\mathrm{eff}}_1=P_0H_0P_0+\alpha P_0(h_{4,5}+h_{5,6}+h_{6,4}-6)P_0+\mathcal{O}(\alpha^{\frac{3}{2}}),
	\label{eq:effH1}
\end{equation}
where $P_0=|p_0\rangle\langle p_0|\otimes\mathbf{1}_{456}$ projects onto the ground state subspace of $H_0$. The ground state of this effective Hamiltonian is
\begin{equation}
	|p_1\rangle=|\mathcal{A}\rangle_{123}\otimes|\mathcal{A}\rangle_{456}.
\end{equation}
For a $N$-layer system in the lattice shown in Fig.~\ref{fig:flower1} with exponentially decaying $J_n=\sqrt{6}\alpha^{n-\frac{1}{2}}$, and Hamiltonian
\begin{equation}
\begin{split}
	H_N=H_0+\sum_{n=1}^{N-1} J_n\big(h_{3n,3n+1}+h_{3n+1,3n-1}+h_{3n,3n+2}\\ +h_{3n+2,3n-2}+h_{3n+3,3n-1}+h_{3n+3,3n-2}\big),
\end{split}
	\label{eq:3Ham}
\end{equation}
 iterating the RG procedure recursively gives the effective Hamiltonian for a lattice with $N$ layers
\begin{equation}
\begin{split}
	H^{\mathrm{eff}}_N=P_{N-1}\sum_{n=0}^{N-1}\alpha^n \left(h_{3n+1,3n+2}+h_{3n+2,3(n+1)}+h_{3(n+1),3n+1}-6\alpha\right)P_{N-1}\\ +\mathcal{O}(\alpha^{N-\frac{1}{2}}),
 \end{split}
\end{equation}
where $P_{n}=|p_n\rangle\langle p_n|$, with
\begin{equation}
	|p_n\rangle=\bigotimes_{m=1}^n|\mathcal{A}\rangle_{3m-2,3m-1,3m},\label{eq:GS}
\end{equation}
projects onto the ground state subspace of $H^{\mathrm{eff}}_n$. The ground state $|p_N\rangle$ is a maximally entangled state in two dimensional space generalizing the rainbow state. When cut along the purple dashed lines in Fig.~\ref{fig:curvature} separating lattice sites into two subsystems, lattice sites of the same radius are entangled within each other giving contributions in units of $\log 3$ each to the EE, making the total EE between even and odd subsystems $N\log 3$. An explicit calculation of the bipartite EE is detailed below.

The Schmidt decomposition of the ground state \eqref{eq:GS} can be written as 
\begin{equation}
	\ket{GS}=\bigotimes_{n=1}^N\left(\frac{1}{\sqrt{3}}\sum_{c_{n}}\ket{c_n}_{3n-1}\otimes\ket{\bar{c}_n}_{3n-2,3n}\right)
\end{equation}
with the normalized wavefunction of the even subsystem
\begin{equation}
	\ket{\bar{c}_n}=\frac{1}{\sqrt{2}}\sum_{a,b=R,G,B}\epsilon_{abc_n}\ket{a}_{3n-2}\otimes\ket{b}_{3n},
\end{equation}
where $\epsilon_{abc_n}$ is the Levi-Civita symbol. Taking the partial trace over the subsystem of even sites, we get the reduced density matrix
\begin{equation}
	\rho_o\equiv\tr_e\ket{GS}\bra{GS}=\frac{1}{3^N}\sum_{c_{n}}\bigotimes_{n=1}^N\ket{c_n}_{3n-1}\bra{c_n}_{3n-1}
\end{equation}
with the constant Schmidt coefficient $1/3^N$. Hence the bipartite entanglement entropy between even and odd subsystems is 
\begin{equation}
	S_N=\sum_{\{c_n\}=R,G,B}\frac{1}{3^N}\log\frac{1}{3^N}=N\log 3.
\end{equation}	

\section{Strong disorder renormalization group with Schrieffer-Wolff transformation}
\label{sec:SW}

The analysis in \ref{sec:Young} concludes that Hilbert space of $H_0$ is fragmented into Krylov subspaces of fixed color content. There is a unique 6-dimensional sector with all three colors that contain the antisymmetrized ground state, six  3-dimensional sectors with two colors, and three 1-dimensional sectors with all three sites in the same color. Only the antisymmetric subspace in the sectors involving only two colors are relevant for the second order perturbation as the ground state $|p_0\rangle$ is antisymmetric and the action of $H_0$ changes only one color among the three sites.

To evaluate the non-vanishing off-diagonal elements of $\langle c'_4c'_5c'_6|\otimes \langle q_k|V_1|p_0\rangle\otimes |c_4c_5c_6\rangle$ for excited state $|q_k\rangle$ of $H_0$ consisting of two different colors and one of $c_4$, $c_5$, or $c_6$, we just need to compute for instance
\begin{align}
	h_{3,4}|p_0\rangle\otimes |R\rangle_4=&\frac{1}{\sqrt{6}}(|RGR\rangle-|GRR\rangle)\otimes|B\rangle_4 \nonumber \\ &+\frac{1}{\sqrt{6}}(|BRR\rangle-|RBR\rangle)\otimes|G\rangle_4,\\
	h_{4,2}|p_0\rangle\otimes |R\rangle_4=&\frac{1}{\sqrt{6}}(|GRR\rangle-|RRG\rangle)\otimes|B\rangle_4 \nonumber \\ &+\frac{1}{\sqrt{6}}(|RRB\rangle-|BRR\rangle)\otimes|G\rangle_4,
\end{align}
so that
\begin{equation}
	V_1|p_0\rangle\otimes |R\rangle_4=\frac{1}{\sqrt{6}}(|RGR\rangle-|RRG\rangle)\otimes|B\rangle_4+\frac{1}{\sqrt{6}}(|RRB\rangle-|RBR\rangle)\otimes|G\rangle_4,
	\label{eq:v1explicit}
\end{equation}
with both of the excited states of $H_0$ of energy $-1$.
To get the diagonal elements of $H_1^\mathrm{eff}$, the color on site 4 must be changed back to $R$, with another action of $h_{3,4}$ or $h_{4,2}$, by the Hermiticity of which,
\begin{equation}
\begin{aligned}
	&\langle R|_4\otimes\langle p_0|V_1\frac{1}{\sqrt{2}}(|RGR\rangle-|RRG\rangle)\otimes|B\rangle_4\\ \equiv &\langle B|_4\otimes\frac{1}{\sqrt{2}}(\langle RGR|-\langle RRG|)V_1|p_0\rangle\otimes|R\rangle_4=\frac{1}{\sqrt{3}}.
\end{aligned} 
\end{equation}
The off-diagonal element that switches the color between site 4 and 5 is obtained by evaluating, for example,
\begin{equation}
\begin{aligned}
	&(h_{1,5}+h_{5,3})\frac{1}{\sqrt{2}}(|RGR\rangle-|RRG\rangle)\otimes|BB\rangle_{45}\\ =&\frac{1}{\sqrt{2}}(|BGR\rangle-|BRG\rangle+|RGB\rangle)\otimes|BR\rangle_{45}-\frac{1}{\sqrt{2}}|RRBBG\rangle.
\end{aligned}
\end{equation}
Hence,
\begin{equation}
	\langle BR|_{45}\otimes\langle p_0|V_1|\frac{1}{\sqrt{2}}(|RGR\rangle-|RRG\rangle)\otimes|BB\rangle_{45}=-\frac{1}{2\sqrt{3}}.
\end{equation}
Due to the $S_3$ symmetry of the model, the effective Hamiltonian after the renormalization in the Shrieffer-Wolff transformation~\cite{PhysRev.149.491,Bravyi:2011aa} is given by
\begin{equation}
\begin{aligned}
	&\langle c_4c_5c_6|\otimes\langle p_0|V_1^\mathrm{eff}|p_0\rangle\otimes|c_4c_5c_6\rangle\\ =&\sum_{i=4}^6\sum_{c'_i\ne c_i}\frac{\langle c_i|\otimes\langle p_0|V_1|q_k\rangle\otimes|c'_i\rangle\langle c'_i|\otimes\langle q_k|V_1|p_0\rangle\otimes|c_i\rangle }{E_0-E_k}=-1,
\end{aligned}
\end{equation}
for the $3^3=27$ diagonal elements; and
\begin{equation}
\begin{aligned}
	&\langle c' c|_{ij}\otimes\langle p_0|V_1^\mathrm{eff}|p_0\rangle\otimes|cc'\rangle_{ij}\\ =&\sum_{c''=c,c'}\frac{\langle c' c|_{ij}\otimes\langle p_0|V_1|q_k\rangle\otimes|c''c''\rangle_{ij}\langle c''c''|_{ij}\otimes\langle q_k|V_1|p_0\rangle\otimes|cc'\rangle_{ij}}{E_0-E_k}=\frac{1}{6},
\end{aligned}
\end{equation}
for the 9 non-vanishing off-diagonal elements.
The $SU(3)$ structure modifies the Dasgupta-Ma rule of the renormalized coupling strength to $\tilde{J}_1=\frac{J_1^2}{6J_0}=\alpha$, giving the effective Hamiltonian \eqref{eq:effH1}.

\section{Lattice geometry and EE scaling}
\label{sec:curve}

\begin{figure}
	\centering
    \begin{subfigure}[t]{0.4\textwidth}
        \centering
        \includegraphics[width=0.8\linewidth]{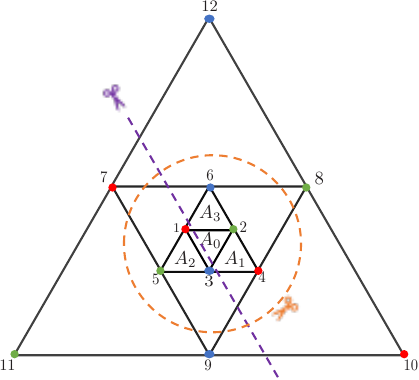} 
        \caption{} \label{fig:curvature1}
    \end{subfigure}
    \begin{subfigure}[t]{0.4\textwidth}
        \centering
        \includegraphics[width=0.8\linewidth]{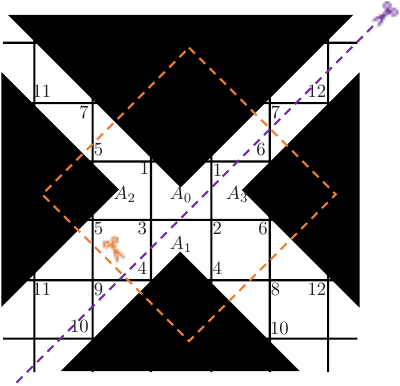} 
        \caption{} \label{fig:curvature2}
    \end{subfigure}
    \begin{subfigure}[t]{0.4\textwidth}
    \centering
        \includegraphics[width=\linewidth]{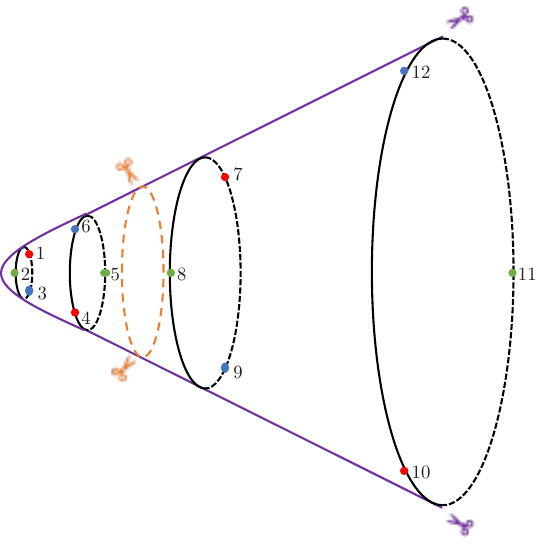} 
        \caption{} \label{fig:curvature3}
    \end{subfigure}
	\caption{(a) The three faces at the center $A_{0,1,2,3}$ have three edges and introduce positive curvature. (b) The lattice can be obtained from a square lattice by cutting out the shaded regions and identifying the lattice sites along the two rays of the same cut-out region. (c) The lattice put on a conical manifold with a smooth apex, in which the interaction strength $J_i$ decays exponentially with the distance on the manifold. The lattice bonds are geodesics between vertices. The purple lines mark the cut between two maximally entangled subsystems, while the perpendicular yellow dashed lines separate minimally entangled subsystems.}
	\label{fig:curvature}
\end{figure}

Before discussing the scaling of the ground state EE, a closer examination of the geometry of the lattice is in order. Since the number of lattice sites grows as $3N$ with the number of layers $N$, the Hausdorff dimension of the lattice is one. Next observe that the interactions are short ranged, as the coupling strength decays exponentially with the distance between neighboring sites, with the exception of the innermost layer. The exception can be eliminated by putting the lattice on a conical surface with a smooth apex as in Fig.~\ref{fig:curvature3}. Another way to look at this is to notice that the vertices in the lattice are all quad-valent. Hence the lattice can be obtained from a square lattice with four disclination defects, which introduces positive curvature only in the four central faces, see Fig.~\ref{fig:curvature2}. The apex angle of the cone decreases as the parameter $\alpha$ gets smaller, in such way that the coupling strength between neighboring sites among the first $6$ lattice sites decays exponentially with the distance along the geodesics of the surface, with $\alpha\to\frac{1}{6}$ corresponding to the flat surface where the perturbation theory fails. Outside the two innermost layers, the spatial lattice is flat, but an nontrivial spacetime metric can nonetheless be derived along the lines of Ref.~\cite{MacCormack:2019aa}, as the spatial distance doubles in each layer.

In this two dimensional manifold where the lattice reside, there are two ways to divide the system that give completely different EE scaling behavior. A cut along the angular direction in Fig.~\ref{fig:curvature1}, or equivalently with a conic cross-section in Fig.~\ref{fig:curvature3}, which separates the system into two concentric subsystems gives zero EE, as the ground state is a product state of each layer. A cut along any of the radial direction in Fig.~\ref{fig:curvature1} or with a triangular cross-section gives an EE that scales with the number of lattice sites in each subsystem. Notice that despite the number of bonds between neighboring site also scales linearly with the subsystem size, this is different from the area law of EE, as the coupling strength forms a geometric series accounting for effectively a bounded constant number of bonds across the cut.

\section{Higher dimensional generaliztion to nested k-simplices}
\label{sec:matryoshka}

\begin{figure}[t!bh]
	\centering
    \begin{subfigure}[t]{0.4\textwidth}
        \centering
        \includegraphics[width=0.4\linewidth]{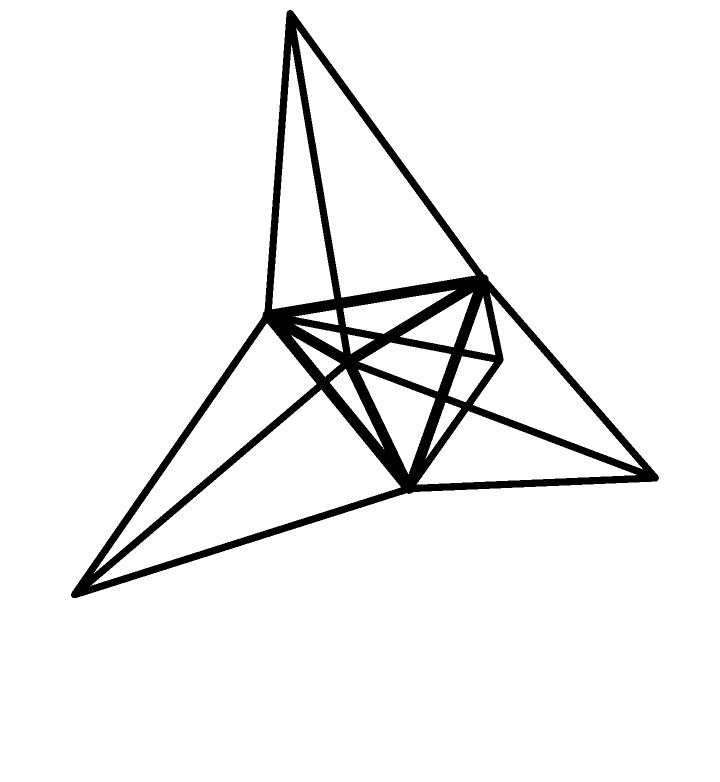} 
        \caption{} \label{fig:4simplices1}
    \end{subfigure} 
    \begin{subfigure}[b]{0.4\textwidth}
        \centering
        \includegraphics[width=0.4\linewidth]{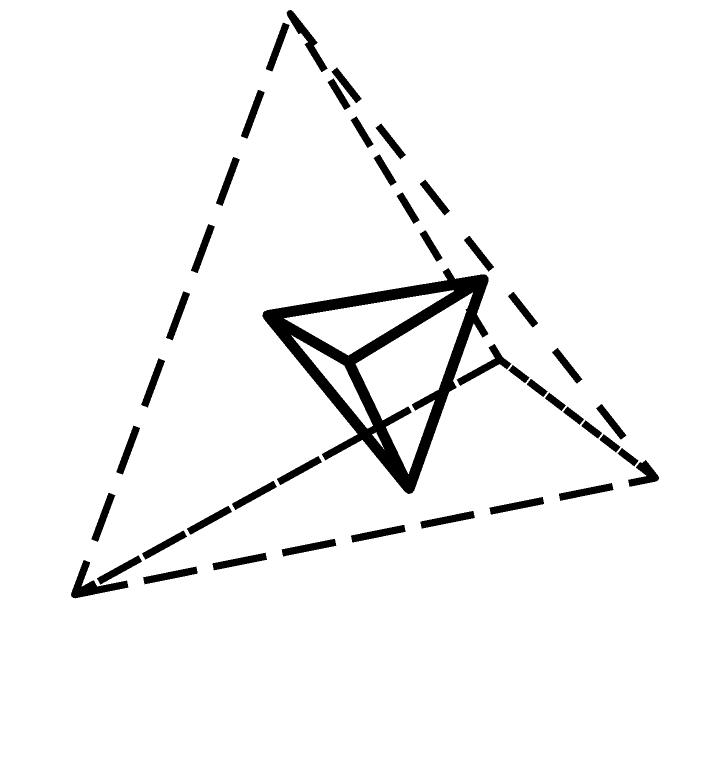} 
        \caption{} \label{fig:4simplices2}
    \end{subfigure}
    
    \begin{subfigure}[t]{0.4\textwidth}
    \centering
        \includegraphics[width=\linewidth]{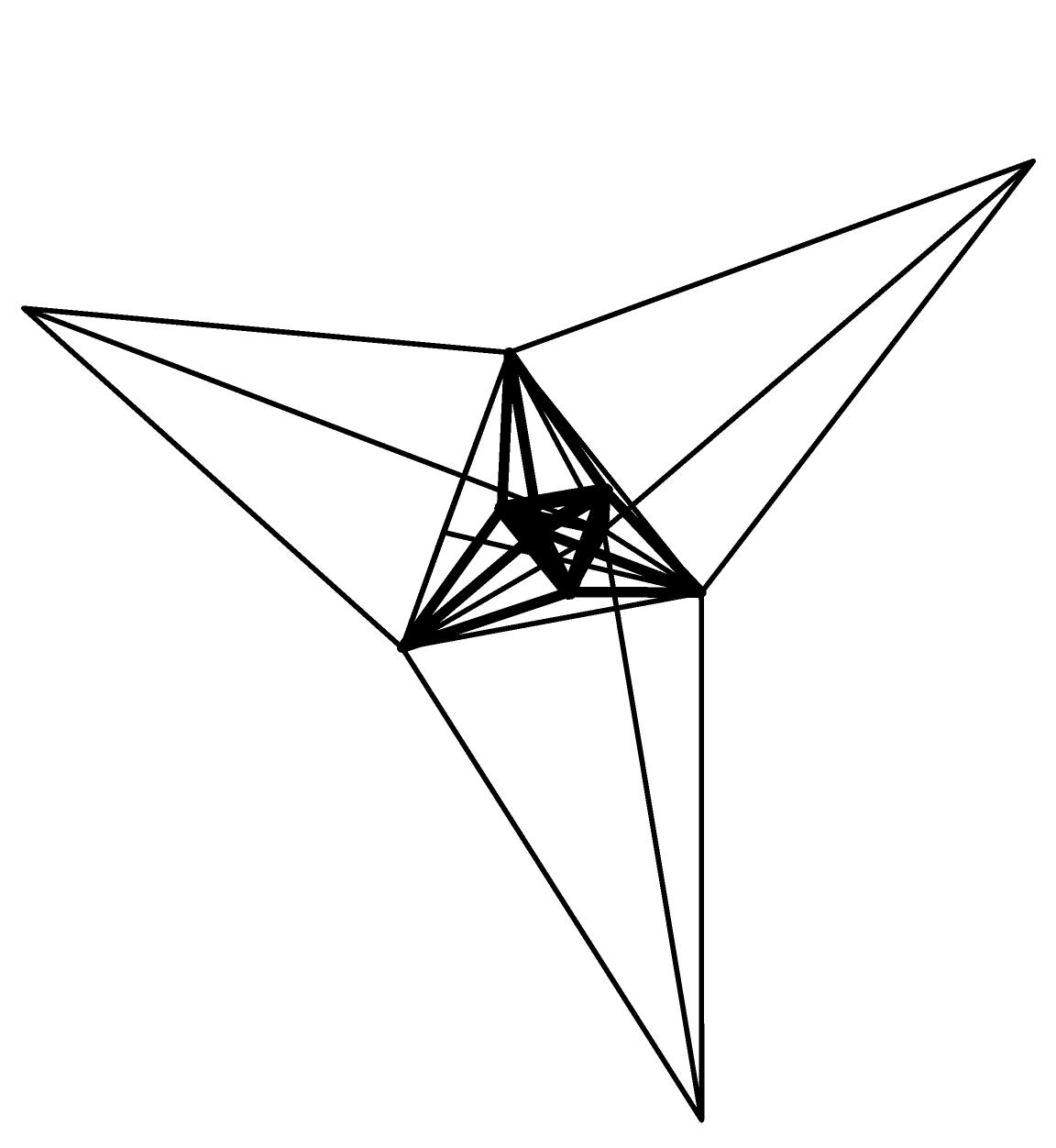} 
        \caption{} \label{fig:4simplices3}
    \end{subfigure}
    \begin{subfigure}[t]{0.4\textwidth}
    \centering
        \includegraphics[width=\linewidth]{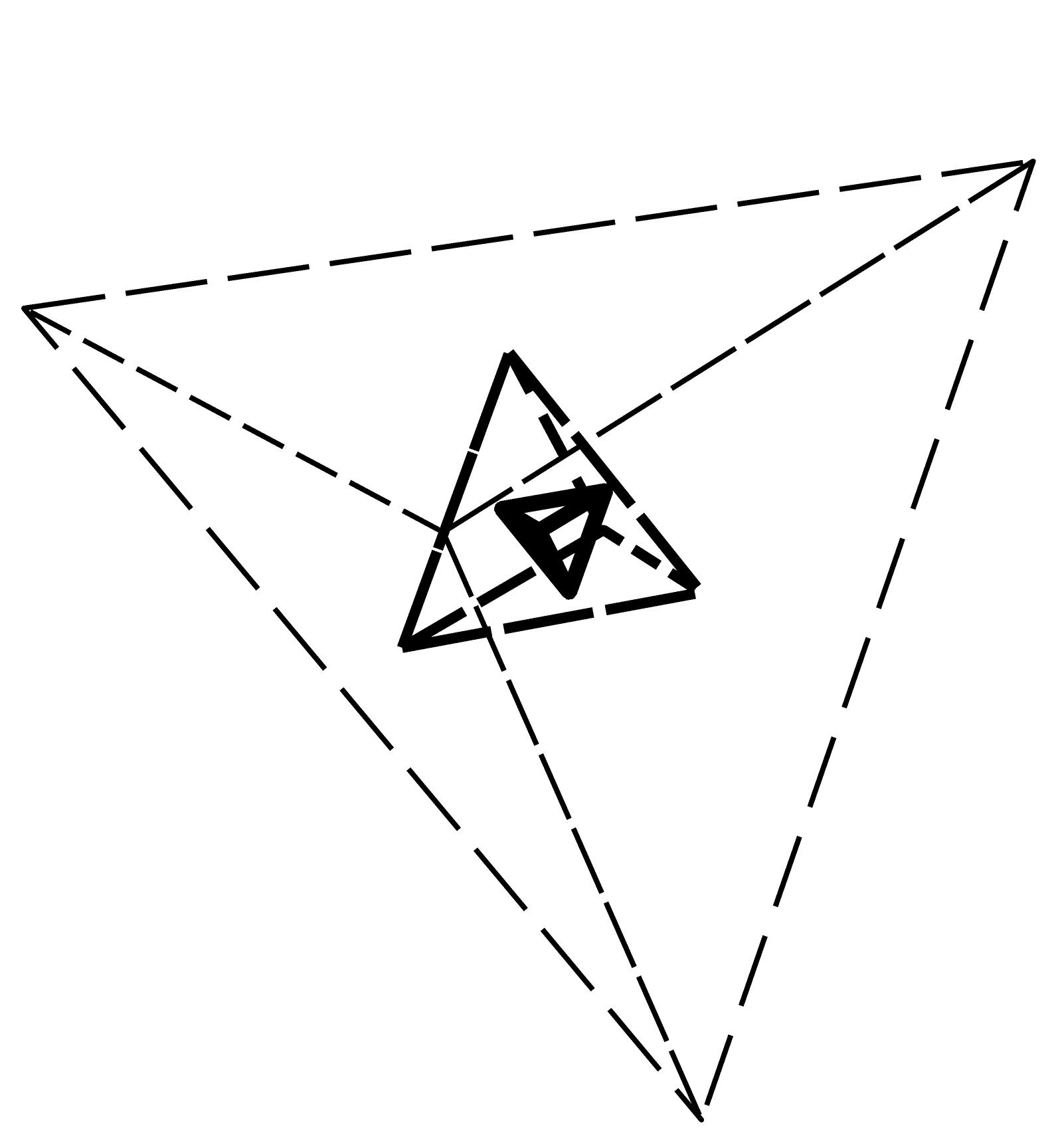} 
        \caption{} \label{fig:4simplices4}
    \end{subfigure}
	\caption{(a) Two layers of nested tetrahedron (3-simplex) lattice with stronger coupling strength in the inner layer than the outer, indicated by thickness of the bonds. (b) The effective longer-range coupling between lattice cites with the same radius. (c) and (d) include the next layer of the inhomogeneous lattice where coupling strength between neighboring sites decays geometrically with distance. In (b) and (d) the vertices of the tetrahedron in an inner layer are located at the face centers of the tetrahedron in the next layer, implying that the linear size of the tetrahedrons triples from one layer to the next.}
	\label{fig:4simplices}
\end{figure}

The Hamiltonian \eqref{eq:3Ham} can be generalized to be $S_{k+1}$ symmetric in the case of nested k-simplex lattice in Fig.~\ref{fig:4simplices1} and \ref{fig:4simplices3}, with sites labeled by linearly increasing integers from $1$ to $kN$ in an $N$-layer system. Each layer consists of $k+1$ lattice sites, each of which span a $k$-simplex with $k$ equidistant neighboring lattice sites from the previous layer. The radius of the new lattice site is chosen such that the lattices sites from the previous layer reside on the plane and precisely at the geometrical center of the $(k-1)$-simplex spanned by the new lattices sites on the same side, as can be seen from Fig.~\ref{fig:4simplices2} and \ref{fig:4simplices4}. This implies that the distance between neighboring sites in a $k$-simplex matryoshka lattice multiplies $k$-fold from one layer to the next, for any $k$.
With the exception of the first layer, as expected due to the curvature at the center like the rainbow chain and the lattice of the previous section, the coupling strength also decays exponentially in proportion to the $k$-dimesnional distance between the lattice sites.

The Hamiltonian can be written as
\begin{equation}
	H_N^{(k)}=\sum_{\substack{i,j=1\\ i\ne j}}^{k+1} h_{i,j}^{(k)} +\sqrt{(k+1)!}\sum_{n=1}^{N-1}\alpha^{\frac{2n-1}{2}}\sum_{\substack{<i,j>\\ i\in [(k+1)(n-1)+1,(k+1)n]\\ j\in [(k+1)n+1,(k+1)(n+1)]}}h_{i,j}^{(k)},
\end{equation}
 where $<i,j>$ denotes a pair of nearest neighboring sites $i$ and $j$, and 
 \begin{equation}
 	h_{i,j}^{(k)}=\sum_{\substack{a,b=1\\ a\ne b}}^{k+1} e^{ab}_ie^{ba}_j,
 \end{equation}
with $(e^{ab})_{cd}=\delta_c^a\delta_d^b$ being the standard basis of $(k+1)\times (k+1)$ matrices.

Following the same renormalization procedure as in Sec.~\ref{sec:flower} and \ref{sec:SW}, as outlined in \ref{sec:diagramatics}, in the $\alpha\ll 1$ limit, it becomes the effective Hamiltonian 
\begin{equation}
	\tilde{H}_N^{(k)}=P^{(k)}_{N-1}\sum_{n=0}^{N-1} \alpha^n \sum_{\substack{i,j=n(k+1)+1\\ i \ne j}}^{(n+1)(k+1)} \left(h^{(k)}_{i,j}-(k+1)!\alpha\right)P^{(k)}_{N-1}+\mathcal{O}(\alpha^{N+\frac{1}{2}}),
\end{equation}
where $P^{(k)}_n=\ket{p^{(k)}_n}\bra{p^{(k)}_n}$, and
\begin{equation}
	\ket{p_n^{(k)}}=\bigotimes_{m=0}^{n-1}\left(\frac{1}{\sqrt{(k+1)!}}\sum_{\sigma\in S_{k+1}}\mathrm{sgn}(\sigma) \bigotimes_{a=1}^{k+1} \ket{c_{\sigma a}}_ {(k+1)m+a}\right).
\end{equation}
The ground state of this effective Hamiltonian is $\ket{p_N^{(k)}}$, and it has similar EE scaling behavior of $N\log(k+1)$ as the two dimensional model, when the two subsystems are divided by a $(k-1)$-dimensional hyperplane that passes through the center of the lattice.

%
\section{Conclusions}
\label{sec:Concl}
%
In this paper, I reported a higher dimensional generalization to the highly entangled rainbow chain, with spatially decaying density of lattice sites in the radial direction, naturally revealing the AdS spacetime metric therein. The dimensionality of the local Hilbert space grows with the dimensionality of the spatial lattice, in order for the effective Hamiltonian to remain frustration free at each order of the perturbation theory. However, it would be interesting to see how much of the result rely on the local degrees of freedom, either with analytical or numerical approaches. 

Some aspects of the model bear resemblances with the SYK model~\cite{PhysRevLett.70.3339, kitaev_talk}, namely the large degrees of freedom and the all-to-all interaction at each layer of the lattice. To some extent, it provides an intermediate setting both between a uniform and a random coupling strength, and between a lattice model with local interaction only between neighbors and a zero-dimensional model where every site interact with each other. At the moment it is not clear how such a model might be useful for understanding black hole physics or metal without quasiparticles. However, recently there have also been similar models constructed more specifically towards realizing AdS/CFT duality on a discrete holographic lattice~\cite{10.21468/SciPostPhys.13.5.103,basteiro2022aperiodic}.

Although one can easily perform a Jordan-Wigner transformation on the lattices of the models discussed here, the outcome does not satisfy a fermionic anti-commutation relation, nor a parafermionic one~\cite{Fendley_2014}. Therefore, much of the recent finding in the rainbow chain about depletion away from half-filling does not apply here directly~\cite{PhysRevB.106.224204}. However, one might be able to construct from scratch another model with similar features with multiple flavors of free fermions with correlated hopping interaction.

Finally, other variants and deformations of the lattice could also be worth exploring. For instance, one can define the same Hamiltonian on the dual lattice of Fig.~\ref{fig:flower} (a) and Fig.~\ref{fig:4simplices} (a), which generalizes a version of the rainbow chain symmetric about a lattice site instead of a bond, as was also discussed in Ref.~\cite{10.21468/SciPostPhys.13.5.103}. The construction of the current models is based on the heuristic picture that when each site on the next layer is adjacent to all but one sites in the previous layer, the state there tends to be the same that non-adjacent (opposite) site, so that sites in each layer forms a singlet by tending to be the same as a different site in the previous layer. So the entanglement pattern is not necessarily the same on an inhomogeneous square lattice or one with a negative curvature resulting from a pentagonal defects. But nonetheless it could be an interesting direction for future studies. Recent deformations to the rainbow chain with inhomogeneous magnetic field~\cite{byles2023q} can also be generalized to the higher dimensional models that allows quantitative control over the entanglement.

\section*{Acknowledgments}
I thank Israel Klich, Bego\~{n}a Mula, Giuseppe Mussardo, Javier Rodr\'{i}guez-Laguna, Germ\'{a}n Sierra and Erik Tonni for fruitful discussions. I gratefully acknowledge support from the Simons Center for Geometry and Physics, Stony Brook University at which some of the research for this paper was performed.

\bibliographystyle{apsrev4-2}
\bibliography{Matryoshka}

\begin{appendix}

\section{Diagonalization of $S_{k+1}$ symmetric Hamiltonians of a single $k$-simplex}
\label{sec:Young}

This appendix details the diagonalization the $S_{k+1}$ symmetric Hamiltonian in a $k$-simplex using the example of $k=2$ and $3$. First for a Hamiltonian written as a sum of permutation, or more specifically, transposition operators, and then show how removing the diagonal terms in the Cartan subalgebra modifies the spectrum without changing the eigenvectors.

Since the Hamiltonian does not change the total number of sites in each configuration, but simply rearranges them, the Krylov subspaces of the fragmented Hilbert space are spanned by elements of the permutation group $S_{k+1}$ acting on one particular configuration with a fixed ordering. Thus an arbitrary choice of the product configuration maps the Krylov subspace to the group algebra of $S_{k+1}$, and the eigenstates of the Hamiltonian, which is an element of the group algebra, becomes the resolution of the unit element into primitive idempotents into minimal left ideals~\cite{hamermesh1989group}. This is the same problem as finding the wave-functions of identical particles of a certain symmetry type and the minimal left ideals are precisely the irreducible representations of the group algebra given by Young operators corresponding to Young Tableaux. The Hamiltonian being the sum of all the group elements of the same conjugacy class, namely those of the cycle type of a single transposition, commutes with all of the group elements, as a result of the rearrangement theorem. Therefore within each irreducible representation of the symmetric group, it has the same eigenvalue.

The Young operator corresponding to a Young tableau $\tau$ is defined as $\mathcal{Y}_\tau=\mathcal{Q_\tau P_\tau}$, where the symmetrizer $\mathcal{P_\tau}$ is the product of sums of all the permutations within each row of $\tau$ and the antisymmetrizer $\mathcal{Q_\tau}$ is the product of algebraic sums of all the permutations within each column of $\tau$ with a sign from the parity of the permutation. The outcome of acting the Hamiltonian on $\mathcal{Y}_\tau$ can be evaluated separately in three parts. First, for terms acting on sites in the same column $c$ of $\tau$, 
\begin{align*}
	\sum_{i\ne j \in c}h_{i,j}\mathcal{Y}_\tau=&\prod_{c'\ne c}\mathcal{Q}_{c'}\sum_{i\ne j\in c}(i,j)\sum_{q\in S_c}\mathrm{sgn}(q)q\prod_{r\in \tau}\mathcal{P}_r\\
	=&-\prod_{c'\ne c}\mathcal{Q}_{c'}\sum_{i\ne j\in c}\sum_{q'\in S_c}\mathrm{sgn}(q')q'\prod_{r\in \tau}\mathcal{P}_r\\
	=&-\binom{l_c}{2}\mathcal{Y}_\tau,
\end{align*}
where $(i,j)$ denotes the transposition between site $i$ and $j$, $l_c$ denotes the length of the column $c$, and the substitution $q'=(i,j)\circ q$ is made in the second  line. Likewise , for terms acting on sites in the same row $r$, we have
\begin{align*}
	\sum_{i\ne j \in r}h_{i,j}\mathcal{Y}_\tau=&\prod_{c \in \tau }\mathcal{Q}_{c}\prod_{r'\ne r}\mathcal{P}_{r'}\sum_{i\ne j\in r}(i,j)\sum_{p\in S_r}p\\
	=&\prod_{c\in\tau}\mathcal{Q}_{c}\prod_{r'\ne r}\mathcal{P}_{r'}\sum_{i\ne j\in r}\sum_{p'\in S_r}p'\\
	=&\binom{l_r}{2}\mathcal{Y}_\tau,
\end{align*}
with $l_r$ denoting the length of row $r$, and $p'=(i,j)\circ p$.
The remaining terms can be grouped into sums of transpositions between each site in one column $c$ and a given site in a different column to show that they sum to zero:
\begin{equation*}
	\sum_{j \in c}h_{i,j}\mathcal{Y}_\tau=\prod_{c'\ne c}\mathcal{Q}_{c'}\sum_{j\in c}(i,j)\sum_{q\in S_c}\mathrm{sgn}(q)q\prod_{r\in \tau}\mathcal{P}_r.
\end{equation*}
Since the first sum on the r.h.s. is symmetric with respect to sites $j$ in column $c$, while the second sum is antisymmetric, their product is identically zero. Hence we have shown that 
\begin{equation}
	H^{(k)}\mathcal{Y}_\tau=\frac{1}{2}\left(\sum_{r\in\tau}l_r(l_r-1)-\sum_{c\in \tau}l_c(l_c-1)\right)\mathcal{Y}_\tau.
\end{equation}
Thus $\mathcal{Y}_\tau$ acting on a certain classical configuration gives the eigenstation of the quantum Hamiltonian of symmetry type $\tau$. Since Young tableaux provide a irreducible representation of the symmetry group $S_{k+1}$, we have found all the eigenstates forming a complete basis.

For instance, in the $SU(3)$ model, the ground state is given by the Young tableau \ytableausetup
{boxsize=1em,aligntableaux=center}\begin{ytableau}
	1 \\
	2 \\
	3
\end{ytableau}
. It has energy $-3$ and is unique because there is only one choice of assigning three different colors to the three rows. The excited states corresponding to standard Young tableaux \begin{ytableau}
	1 & 2 \\
	3
\end{ytableau} and \begin{ytableau}
	1 & 3 \\
	2
\end{ytableau} can act nontrivially on configurations with at least colors. In the Krylov subspace that contain three colors, since site $2$ and $3$ are neither symmetrized nor antisymmetrized, there are $2$ different ways to assign colors to the boxes for tableau, namely by acting the corresponding Young operators on the vectors $\ket{RGB}$ and $\ket{RBG}$. And in the two color subspace, there are $3\times 2=6$ choices of coloring for the two rows as they cannot be in the same color. In total they give $16$ degenerate excited states of energy $0$. The highest energy excited state is given by the tableau \begin{ytableau}
	1 & 2 & 3
\end{ytableau}. It has energy $+3$ and can act on any color configuration, giving a total of $1+\binom{3}{2}+3=10$ degenerate states.

In the $SU(4)$ model, the unique ground state is given by the Young tableau \begin{ytableau}
	1\\
	2\\
	3\\
	4
\end{ytableau}, with energy $-6$. The first excited states come from \begin{ytableau}
	1 & 2\\
	3\\
	4
\end{ytableau}, \begin{ytableau}
	1 & 3\\
	2\\
	4
\end{ytableau}, and \begin{ytableau}
	1 & 4\\
	2\\
	3
\end{ytableau}, with energy $-2$ and degeneracy $3\times\left( 3+\times 4 \times \binom{3}{2}\right)= 45$. The second excited states of energy $0$ are from \begin{ytableau}
	1 & 2 \\
	3 & 4
\end{ytableau}, and \begin{ytableau}
	1 & 3 \\
	2 & 4
\end{ytableau}, with degeneracy $2\times \left(3\times 2 +4\times 2+\binom{4}{2}\right)=40$. The third excited states of energy $+2$ come from
\begin{ytableau}
	1 & 2 & 3\\
	4
\end{ytableau}, \begin{ytableau}
	1 & 2 & 4\\
	3
\end{ytableau}, and \begin{ytableau}
	1 & 3 & 4\\
	2
\end{ytableau}, with degeneracy $3\times \left(3+4\times 2\times 3+\binom{4}{2}\times 3\right)=135$. And the highest energy eigenstates correspond to 
\begin{ytableau}
	1 & 2 & 3 & 4
\end{ytableau}, with energy $+6$ and degeneracy $1+4\times 3+\binom{4}{2}+4\times \binom{3}{2} + 4 = 35$.
Note that the eigenstates generated by the Young operators although linearly independent, are not orthonormal. In the next appendix, we use a graphical notation to evaluate the matrix elements of the Hamiltonian in an overcomplete basis of dimers and simplex singlets with fixed eigenvalue.

Now the Hamiltonian terms in \eqref{eq:offdiagh} deviate from the permutation operator by omitting all the diagonal terms from the Cartan subalgebra of $SU(k+1)$. Yet, since all the eigenstates above are superpositions permutations of the same set of configurations, each of which has the same eigenvalue of the difference between the two Hamiltonians
\begin{equation}
	\delta H\vcentcolon= H^{(\mathrm{perm})}-H^{(\mathrm{od})}=\sum_{\substack{i,j=1\\ i\ne j}}^{k=1}\sum_{a=1}^{k+1} e^{aa}_ie^{aa}_j,
\end{equation}
the eigenvectors remain the same after removing the diagonal terms. So we just need to evaluate the eigenvalues of $\delta H$ and modify the eigenvalues accordingly. Since $\delta H$ simply counts the number of pairs in the same state, this correction is just $-\sum_{n_i\ge 2}\binom{n_i}{2}$, for $c_i$ appearing $n_i$ number of times in each configuration. This will spit the degeneracy of eigenstates corresponding to the same Young tableaux but with different color content. But for our purpose of performing the strong disorder renormalization group in the next appendix, the only relevant information is the eigenvalue of eigenstates corresponding to Young diagrams \ydiagram{2,1}\
 and \ydiagram{2,1,1}\
 with only one color appearing twice, now have energies $-1$ and $-3$ respectively, as opposed to the eigenvalues $0$ and $-2$ for the Hamiltonian with diagonal terms included.

\section{Diagramatics of overcomplete k-simplex singlet basis}
\label{sec:diagramatics}

The explicit evaluation of matrix elements in Sec.~\ref{sec:SW} can be simplified with a graphical notation. Assigning an ordering to the color degree of freedom $c_1<c_2<\cdots$, we can denote the antisymmetrized singlet state between two neighboring sites by a directed bond
\begin{equation}
	\ket{\twosig}_{ij}=|c_1c_2\rangle_{ij}-|c_2c_1\rangle_{ij},
\end{equation}
where the sign is positive when the color indices increase following the direction of the arrow and negative when against, such that
\begin{equation}
	\ket{\ntwosig}_{ij}\equiv-\ket{\twosig}_{ij}.
\end{equation}
Due to the overcompleteness of the singlet basis, we have
\begin{equation}
	\ket{\AB}_{\substack{j\\i\ k}}+\ket{\BC}_{\substack{j\\i\ k}}\equiv\ket{\AC}_{\substack{j\\i\ k}}.
\end{equation}
Further define
\begin{equation}
	\ket{\threesig}_{\substack{j\\i\ k}}\equiv -\ket{\nthreesig}_{\substack{j\\i\ k}}=\sum_{\sigma\in S_3}\mathrm{sgn}(\sigma) \ket{c_{\sigma 1}}_i\otimes\ket{c_{\sigma 2}}_j\otimes\ket{c_{\sigma 3}}_k,
\end{equation}
where the sign is now determined by the signature of the permutation of the ordering of color indices along the direction of the arrow.

The sum over Hamiltonian terms acting on a $(k-1)$-simplex face of a $k$-simplex and an external site in the $SU(k)$ model annihilates all the same color as the $k+1$'th site, leading to a rewriting of \eqref{eq:v1explicit} in the $k=3$ case
\begin{equation}
	(h_{j,l}+h_{k,l})\ket{\foursite}_{\substack{\ j\ l\\i\ k} }=\ket{\foursiteA}_{\substack{\ j\ l\\i\ k}}+\ket{\foursiteB}_{\substack{\ j\ l\\i\ k}}.
\end{equation}
The r.h.s. overlaps only with the excited state corresponding to the Young diagram \ydiagram{2,1}\ as there a only two colors out of three involved among sites $i,j,k$ and one pair is antisymmetrized. Likewise, we can denote the singlet ground state for $k=4$ as
\begin{equation}
\begin{aligned}
	\ket{\foursig}_{\substack{j\\ \\ l\\i\ \ \ \ k}}&\equiv -\ket{\nfoursig}_{\substack{j\\ \\ l\\i\ \ \ \ k}}\\ &=\sum_{\sigma\in S_4}\mathrm{sgn}(\sigma) \ket{c_{\sigma 1}}_i\otimes\ket{c_{\sigma 2}}_j\otimes\ket{c_{\sigma 3}}_k\otimes\ket{c_{\sigma 4}}_l.
\end{aligned}
\end{equation}
Then the analogous relation need for the perturbation calculation becomes
\begin{equation}
\begin{split}
	&(h_{j,m}+h_{k,m}+h_{l,m})\ket{\fivesite}_{\substack{j\\ \ \ \ \ \ m\\ l\\i\ \ \ \ k}}\\
	=&\ket{\fivesiteA}_{\substack{j\\ \ \ \ \ \ m\\ l\\i\ \ \ \ k}}+\ket{\fivesiteB}_{\substack{j\\ \ \ \ \ \ m\\ l\\i\ \ \ \ k}}+\ket{\fivesiteC}_{\substack{j\\ \ \ \ \ \ m\\ l\\i\ \ \ \ k}}.
\end{split}
\end{equation}
Once again, the r.h.s. overlaps only with the excited state corresponding to the Young diagram \ydiagram{2,1,1}\ as there a only three colors out of four involved among site $i,j,k,l$ and one triplet is antisymmetrized. 

The diagramatics introduced in this appendix generalizes to higher simplices without much difficulty and bears resemblance to the over-completeness of dimer basis used, for example, in the solution of exact excited states of the AKLT model~\cite{PhysRevB.98.235155} in one dimension, and the homological method recently employed to study entanglement structure~\cite{PhysRevB.107.115174}. It would be interesting to see if these techniques prove useful in a more generic setting.

\end{appendix}

\end{document}